\documentclass[letter]{aa}
\usepackage{keyval}
\usepackage[english]{babel}
\usepackage{graphicx}
\usepackage{txfonts}
\usepackage{enumitem}
\usepackage{subcaption}
\usepackage{lscape}
\usepackage{placeins}
\usepackage{hyperref}

\begin{document}

\title{Discovery of a nova super-remnant surrounding the recurrent nova LMCN 1971-08a in the Large Magellanic Cloud}

\author{Michael W. Healy-Kalesh\inst{1,2}\fnmsep\thanks{Corresponding author: michael.healy-kalesh@fau.de or M.W.HealyKalesh@ljmu.ac.uk. Honorary Visiting Research Fellow at Liverpool John Moores University.}
        \and Manami Sasaki\inst{1}
        \and Sean D. Points\inst{3}
        \and Miroslav D. Filipović\inst{4}
        \and Zachary J. Smeaton\inst{4}
        \and \\ Matthew J. Darnley\inst{2}
        \and Knox S. Long\inst{5,6}
        \and Sara Saeedi\inst{1}
        \and Federico Zangrandi\inst{1}
        }
\institute{Dr. Karl Remeis Observatory, Erlangen Centre for Astroparticle Physics, Friedrich-Alexander-Universität Erlangen-Nürnberg, \\ Sternwartstraße 7, 96049 Bamberg, Germany
\and Astrophysics Research Institute, Liverpool John Moores University, Liverpool, L3 5RF, UK 
\and NSF’s NOIRLab/CTIO, Casilla 603, La Serena, 1700000, Chile
\and School of Science, Western Sydney University, Locked Bag 1797, Penrith, NSW 2751, Australia
\and Space Telescope Science Institute, 3700 San Martin Drive, Baltimore, MD 21218, USA
\and Eureka Scientific, Inc. 2452 Delmer Street, Suite 100, Oakland, CA 94602-3017, USA}

\date{Received 30 July 2025 / Accepted 16 September 2025}
\abstract
{A nova super-remnant (NSR) is a greatly-extended structure grown by repeated nova eruptions sweeping the surrounding material away from a nova into a dense outer shell and are predicted to form around all novae. To date, four NSRs are known, with three located in the Galaxy and one residing in M31. Here we present the discovery of the first NSR in the Large Magellanic Cloud and only the second extragalactic nova shell identified, hosted by the recurrent nova LMCN 1971-08a. The structure is coincident with the nova, has a circular morphology, and is visible in narrowband H$\alpha$ and $[\ion{S}{ii}]$ filters but is very faint in $[\ion{O}{iii}]$, as expected. $\ion{\text{H}}{i}$ data also potentially reveal the existence of a coincident structure. Further, with a diameter of ${\sim}200$ pc, this NSR is the largest example yet found, with models indicating an ${\sim}4130 \ \text{M}_{\odot}$ shell expanding at ${\sim}20 \ \text{km} \ \text{s}^{-1}$ into the surrounding medium and an age of $\sim$2.4 Myr. The existence of the NSR also suggests that LMCN 1971-08a may have a much shorter recurrence period than currently presumed.}

\keywords{Stars: novae, cataclysmic variables -- Stars: individual: LMCN 1971-08a -- Magellanic Clouds -- ISM: structure -- Radio continuum: ISM}

\authorrunning{Healy-Kalesh et al.}
\titlerunning{Nova super-remnant in the LMC}

\maketitle

\section{Introduction}\label{Introduction}
\begin{figure*}[h!]
\centering
\includegraphics[width=\textwidth]{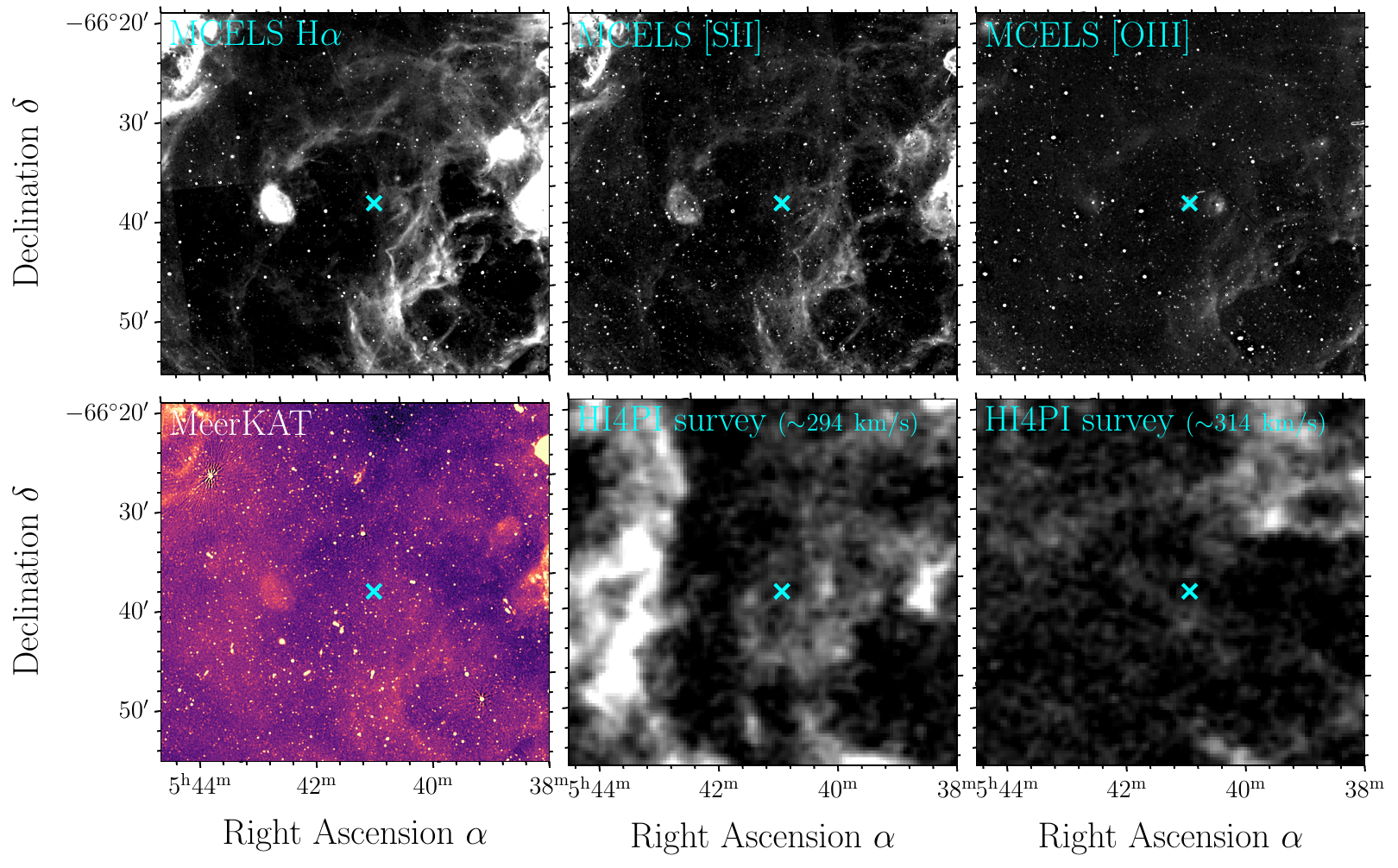}
\caption{Approximate $40 \times 40$ arcminute field of view showing the nova super-remnant around LMCN 1971-08a seen across different surveys and wavebands. The location of the nova is indicated with the cyan cross in each panel. The top row shows the surroundings of the nova in MCELS continuum-subtracted H$\alpha$, $[\ion{\text{S}}{ii}]$ and $[\ion{\text{O}}{iii}]$ data, with the NSR apparent in H$\alpha$ and $[\ion{\text{S}}{ii}]$ but negligible in $[\ion{\text{O}}{iii}]$. The bright "stars" in the MCELS images are points of oversubtraction from saturated stars. The left panel on the bottom row showing MeerKAT data does not clearly reveal a corresponding structure. The middle and right panels in the bottom row show data cube slices at ${\sim}294 \ \text{km} \ \text{s}^{-1}$ and ${\sim}314 \ \text{km} \ \text{s}^{-1}$, respectively, from the HI4PI survey, with potential evidence for the same northeast (in the $314 \ \text{km} \ \text{s}^{-1}$ panel) and southwest (in the $294 \ \text{km} \ \text{s}^{-1}$ panel) components of the NSR seen in the optical.}
\label{observations}
\end{figure*}

\noindent Recurrent novae (RNe) are observationally-defined as accreting white dwarf (WD) binaries that have exhibited more than one nova eruption. Such eruptions are driven by a thermonuclear runaway on the WD's surface \citep{1972ApJ...176..169S}, which leads to a portion of the accreted material being ejected at velocities up to thousands of kilometers per second \citep{2001IAUS..205..260O}; in some cases, this material forms expanding sub-parsec nova shells \citep{2024MNRAS.529..212S,2025MNRAS.539..246S}. After the eruption, the resumption of accretion onto the WD \citep{2007MNRAS.379.1557W} prepares the nova system for subsequent eruptions. Furthermore, accumulation of material retained on the WD following each eruption can grow the WD \citep{2005ApJ...623..398Y,2016ApJ...819..168H} to the Chandrasekhar limit to potentially explode as a type Ia supernova \citep{1973ApJ...186.1007W}.

Nova super-remnants (NSRs) are shell-like structures orders of magnitude larger than singular eruption nova shells \citep[for a review see][]{2024gacv.confE..35H}. The first NSR \citep{2019Natur.565..460D} was discovered in the Andromeda Galaxy around the rapidly recurring nova, M31N 2008-12a and formed by the ejecta of repeated nova eruptions over millennia, excavating tens of thousands of solar masses of surrounding interstellar medium from around the central binary into a high-density shell \citep{2019Natur.565..460D}. Theoretically, such structures should encompass all novae \citep{2023MNRAS.521.3004H}, although their size on the sky and low surface brightnesses make NSRs difficult to find in optical data. However, with the capabilities of the recently deployed Condor Array Telescope \citep{2023PASP..135a5002L}, optical NSRs have now been discovered around the Galactic RNe KT Eridani \citep{2024MNRAS.529..224S,2024MNRAS.529..236H}, T Coronae Borealis \citep{2024ApJ...977L..48S} and RS Ophiuchi \citep{2025AJ....170...56S}\footnote{A possible cavity component of the NSR associated with RS Ophiuchi was identified in far-infrared data \citep{2024MNRAS.529L.175H}.}. These NSRs have diameters ranging from ${\sim}30-130$ pc; have emission most prominent in H$\alpha$, $[\ion{S}{ii}]$ and $[\ion{N}{ii}]$; lack strong $[\ion{O}{iii}]$ emission as a result of ionisation from low velocity shocks; and exhibit low expansion velocities in their outer shells. While the NSR around M31N 2008-12a appears to have a more well-defined elliptical shell as it is viewed pole-on \citep{2019Natur.565..460D}, the Galactic NSRs have more patchy morphologies likely as a result of the NSR growing within the irregular interstellar medium (ISM) and possibly from being observed at different viewing angles. Furthermore, the NSRs surrounding T CrB \citep{2024ApJ...977L..48S} and RS Oph \citep{2025AJ....170...56S} appear bilobed, possibly resulting from the ejecta being initially influenced by the accretion disk around the WD. A search for additional NSRs in M31 and the Large Magellanic Cloud (LMC) did not yield further examples \citep{2024MNRAS.528.3531H}, but this apparent dearth was attributed to the structures around other "younger" systems being considerably fainter.

LMCN 1971-08a is one of the four currently known recurrent novae in the LMC. It exhibited eruptions in 1971 \citep{1971IAUC.2353....1G} and 2009 \citep{2009IAUC.9019....1L} and therefore has a recurrence period (or inter-eruption interval) of ${\sim}38$/N years \citep{2016ApJ...818..145B}, where N may be one or a small integer. While only a spectrum of the first eruption was obtained in 1971 with no further follow-up \citep{1971IAUC.2353....1G,2016ApJ...818..145B}, the more recent eruption in 2009 was followed in great detail, being observed in the optical, UV, near-infrared \citep{2016ApJ...818..145B}, and X-rays \citep{2016ApJ...818..145B,2021MNRAS.505.3113O}. These observations revealed a very fast-declining nova comprising a $1.1-1.3 \ \text{M}_{\odot}$ WD, accreting material at an elevated rate of $3.6^{+4.7}_{-2.5} \times 10^{-7} \ \text{M}_{\odot} \ \text{yr}^{-1}$ from a subgiant companion with a probable orbital period of 1.2 days \citep{2016ApJ...818..145B}. The ejecta velocities seen from the nova ranged from $1000-4200 \ \text{km} \ \text{s}^{-1}$ \citep{2009ATel.1930....1O,2016ApJ...818..145B} with shocks likely present in the ejecta before day nine of the eruption \citep{2016ApJ...818..145B}.

In this Letter, we present the discovery of a nova super-remnant around the recurrent nova LMCN 1971-08a, the first nova super-remnant to be found in the LMC. In Sect.~\ref{Data} we outline the datasets used in this work before describing the characteristics of the NSR across optical and radio data in Sect.~\ref{Nova super-remnant}. In Sect.~\ref{Discussion} we place the discovery within the context of the growing field of nova super-remnants and rule out other possible origins before summarising our findings in Sect.~\ref{Summary}.

\section{Optical and radio data}\label{Data}
Within this study, several datasets were utilised across the optical and radio. The Magellanic Cloud Emission Line Survey \citep[MCELS;][]{1999IAUS..190...28S} observed the central $8^{\circ} \times 8^{\circ}$ of the LMC with five filters including narrowband H$\alpha$, $[\ion{S}{ii}]$ and $[\ion{O}{iii}]$ filters. The DeMCELS survey \citep{2024ApJ...974...70P} is a contemporary higher-resolution emission-line survey of the Magellanic Clouds undertaken with the Dark Energy Camera \citep[DECam;][]{2015AJ....150..150F} narrowband H$\alpha$ and $[\ion{S}{ii}]$ filters. The continuum-subtracted narrowband data from both MCELS and DeMCELS were utilised for this work. The HI4PI survey \citep{2016A&A...594A.116H} is an all-sky neutral atomic hydrogen ($\ion{\text{H}}{i}$) survey constructed from the Eﬀelsberg-Bonn $\ion{\text{H}}{i}$ Survey \citep[EBHIS;][]{2011AN....332..637K} and the Galactic All-Sky Survey \citep[GASS;][]{2009ApJS..181..398M} data sets, which provide better angular resolution and sensitivity than previous all-sky $\ion{\text{H}}{i}$ surveys. The MeerKAT radio data (Cotton et al., in preparation) used in this work are from the MeerKAT array \citep{2016mks..confE...1J}, a new generation radio telescope with high resolution and sensitivity such that it is capable of resolving structures in the LMC \citep[see, e.g.,][]{2025A&A...693L..15S}.

\section{LMCN 1971-08a nova super-remnant}\label{Nova super-remnant}
A coherent, circular shell-like structure spatially coincident with LMCN 1971-08a appears in the MCELS H$\alpha$ data (see Fig.~\ref{observations}), with the same shell visible in the MCELS $[\ion{S}{ii}]$ image but negligible in MCELS $[\ion{O}{iii}]$, as expected for NSRs \citep{2019Natur.565..460D,2024MNRAS.529..224S,2024ApJ...977L..48S,2025AJ....170...56S}. While clearly evident in the MCELS data, the high resolution DeMCELS H$\alpha$ and $[\ion{S}{ii}]$ imaging of the NSR emphasizes the structure further, as shown in Fig.~\ref{DeMCELS} (with the approximate extent of the NSR indicated).

With a distance of $49.6 \pm 0.6$ kpc to the LMC \citep{2019Natur.567..200P}, the NSR has a projected diameter of $\sim$200 pc. As seen in the optical data of Fig.~\ref{observations}, the NSR is brighter to the northeast and southwest, with a fainter boundary connecting these two components to the northwest, defining the outer shell edge of the NSR. The shell thickness is taken to be ${\sim}14\%$ from the southwest edge, as the inner and outer boundaries of this part of the NSR shell are ${\sim}87$ pc and ${\sim}101$ pc from the nova, respectively. This possible compression front may indicate interaction of a bow shock with an earlier phase of the remnant's evolution, as seen with a similar feature in the KT Eri NSR, \citep{2024MNRAS.529..236H}. Additionally, there appears to be an interaction with a complex structure to the west (Fig.~\ref{full DeMCELS}) composed of many $\ion{\text{H}}{ii}$ regions \citep{2012ApJ...755...40P}.

Following \citet{2024ApJ...974...70P}, the H$\alpha$ and $[\ion{S}{ii}]$ surface brightnesses of the bright northeast component of the NSR are $4.1 \times 10^{-17} \text{ erg} \text{ cm}^{-2} \text{ s}^{-1} \text{ arcsec}^{-2}$ and $2.5 \times 10^{-17} \text{ erg} \text{ cm}^{-2} \text{ s}^{-1} \text{ arcsec}^{-2}$, respectively. The H$\alpha$ and $[\ion{S}{ii}]$ surface brightnesses of the bright southwest component are marginally fainter, radiating at $2.2 \times 10^{-17} \text{ erg} \text{ cm}^{-2} \text{ s}^{-1} \text{ arcsec}^{-2}$ and $1.4 \times 10^{-17} \text{ erg} \text{ cm}^{-2} \text{ s}^{-1} \text{ arcsec}^{-2}$, respectively.

\subsection{NSR in radio data}\label{NSR in radio data}
As shown in Fig.~\ref{observations}, the $\ion{\text{H}}{i}$ data cube from the HI4PI survey \citep{2016A&A...594A.116H} sliced at ${\sim}294 \ \text{km} \ \text{s}^{-1}$ and ${\sim}314 \ \text{km} \ \text{s}^{-1}$, reveals a correlated structure coincident with the H$\alpha$ shell. Specifically, the southwestern portion of the NSR shows a higher level of emission from in the ${\sim}294 \ \text{km} \ \text{s}^{-1}$ slice, while a possible shell boundary appears to the northeast in the ${\sim}314 \ \text{km} \ \text{s}^{-1}$ slice. A velocity distribution across the NSR, spanning $280-320 \ \text{km} \ \text{s}^{-1}$ in declination and right ascension (see Fig.~\ref{HI slits} for the slits used) reveals an arc with a range of ${\sim}20 \ \text{km} \ \text{s}^{-1}$, as shown in Fig.~\ref{PV diagram}, possibly indicating an outer shell expansion velocity of ${\sim}10 \ \text{km} \ \text{s}^{-1}$. In MeerKAT radio data (Cotton et al., in preparation), there does not appear to be a corresponding structure, as shown in Fig.~\ref{observations}. By estimating the RMS noise level in the region of the NSR shell, we derive an upper limit of ${\sim}30 \ \mu$Jy / beam.

\subsection{Simple model of NSR evolution}\label{Simple model of NSR evolution}
To understand the evolutionary history of the NSR, we employed the hydrodynamical code \texttt{Morpheus} \citep{2007ApJ...665..654V} following \citet{2023MNRAS.521.3004H,2024MNRAS.529..236H}, to produce a simple model of the growth of such a remnant with a radius of ${\sim}100$ pc. For the eruption characteristics of LMCN 1971-08a, we utilised the data from \citep{2016ApJ...818..145B}, taking the recurrence period of the nova to be 38 years; the mass loss phase to be 10.4 days (equal to the $V-$band $t_3$ decline time); the ejecta velocity to be $4000 \ \text{km} \ \text{s}^{-1}$; and the ejecta mass to be $1.4 \times 10^{-5} \ \text{M}_{\odot}$. 

To estimate the surrounding ISM density for the model, we took the $\ion{\text{H}}{i}$ data velocity cube from the HI4PI survey \citep{2016A&A...594A.116H}, converted the integrated intensity in the range of $280-320 \ \text{km} \ \text{s}^{-1}$ into column density $N_{\text{H}}$ \citep[using Eq. 2 from][]{2016A&A...594A.116H}, and found the average $N_{\text{H}}$ within $12'$ of the nova (to incorporate the NSR and its surroundings). Assuming a depth of 5 kpc \citep[estimated from the NSR's location in Fig. 9 of][]{2009A&A...496..399S} gives an ISM density of $6.68 \times 10^{-26} \ \text{g} \ \text{cm}^{-3}$ (number density of $n=0.04 \ \text{cm}^{-3}$).

With a nova system mimicking LMCN 1971-08a in a pre-populated ISM estimated for the local region, it would take $2.37 \times 10^6$ years over ${\sim}$62,400 eruptions to reach a radius of 100 pc, containing an evacuated cavity region ${\sim}$15 pc in radius. The NSR outer shell would have a mass of ${\sim}4130 \ \text{M}_{\odot}$ (made up almost exclusively of ISM) and an expansion velocity of ${\sim}20 \ \text{km} \ \text{s}^{-1}$, which is consistent with the expansion velocity derived from the $\ion{\text{H}}{i}$ position-velocity analysis from Sect.~\ref{NSR in radio data}.

\section{Discussion}\label{Discussion}

\subsection{Comparison to other NSRs}\label{Comparison to other NSRs}
With a diameter of ${\sim}200$ pc, the LMCN 1971-08a NSR is $\sim$1.5 times larger than the NSR around M31N 2008-12a \citep[major axis of 134 pc;][]{2019Natur.565..460D} and a factor of $\sim$4, 3, and 6.5 times larger than the NSRs associated with KT Eri \citep[$\sim$50 pc;][]{2024MNRAS.529..224S}, T CrB \citep[$\sim$30 pc;][]{2024ApJ...977L..48S} and RS Oph \citep[$\sim$70 pc;][]{2025AJ....170...56S}, respectively. The projected morphology of the NSR is also more circular than the other NSRs, likely resulting from the system's inclination, and is most similar to the M31N 2008-12a NSR with more defined edges than the irregular morphology of the Galactic NSRs.

As with all other NSRs, this NSR is most prominent through its H$\alpha$ and $[\ion{S}{ii}]$ emission. The H$\alpha$ surface brightnesses of the brightest component of the NSR (northeast part) is $4.1 \times 10^{-17} \text{ erg} \ \text{ cm}^{-2} \text{ s}^{-1} \text{ arcsec}^{-2}$. Having an H$\alpha$ surface brightness comparable to the other NSRs -- $4 \times 10^{-17} \text{ erg} \text{ cm}^{-2} \text{ s}^{-1} \text{ arcsec}^{-2}$ for KT Eri \citep{2024MNRAS.529..224S}, ${\sim}6.6 \times 10^{-18} \text{ erg} \text{ cm}^{-2} \text{ s}^{-1} \text{ arcsec}^{-2}$ for T CrB \citep{2024ApJ...977L..48S}, ${\sim}1.4 \times 10^{-17} \text{ erg} \text{ cm}^{-2} \text{ s}^{-1} \text{ arcsec}^{-2}$ for RS Oph \citep{2025AJ....170...56S}, and ${\sim}1 \times 10^{-16} \text{ erg} \text{ cm}^{-2} \text{ s}^{-1} \text{ arcsec}^{-2}$ for M31N 2008-12a (based on integrated H$\alpha + [\ion{N}{ii}]$ flux; \citealt{2019Natur.565..460D}) -- this NSR is intrinsically brighter than all others, supporting predictions that larger NSRs should be brighter \citep{2023MNRAS.521.3004H}. Moreover, the NSR shell is almost negligible in the $[\ion{O}{iii}]$ narrowband filter (Fig.~\ref{OIII ratio}) due to the lack of photoionisation across the structure, consistent with the findings of other NSRs \citep{2019Natur.565..460D,2024MNRAS.529..224S,2024ApJ...977L..48S,2025AJ....170...56S}. However, the southwest part does exhibit very faint $[\ion{O}{iii}]$ emission coinciding with a higher $[\ion{S}{ii}]$/H$\alpha$ ratio.

The $\ion{\text{H}}{i}$ position-velocity analysis indicates a possible expansion velocity of the NSR outer edges of ${\sim}10 \ \text{km} \ \text{s}^{-1}$. These velocities are consistent with those expansion velocities seen in other NSRs shells \citep{2019Natur.565..460D,2024MNRAS.529..224S,2025AJ....170...56S} and are in agreement with the range of NSR outer shell velocities predicted by modelling in this work and other studies \citep[e.g.,][]{2023MNRAS.521.3004H}.

Both LMCN 1971-08a and another NSR-hosting nova KT Eri exhibit "several remarkable similarities" \citep{2016ApJ...818..145B}. Those authors suggested that both novae may belong to a class of RNe with WDs approximately $0.1-0.3 \ \text{M}_{\odot}$ below the Chandrasekhar limit, but with very high mass transfer rates between eruptions. This similarity may indicate a potential correlation between these types of novae and the brightest NSRs, although this cannot be tested further until the sample of known NSRs increases.

\subsection{Other origins}\label{Other origins}
It is necessary to consider alternative origins of the visible shell encompassing LMCN 1971-08a, including it being a supernova remnant (SNR), a superbubble, a planetary nebula (PN), or an $\ion{\text{H}}{ii}$ region. We therefore searched for sources within the NSR (${\sim}7'$ radius) using SIMBAD \citep{2000A&AS..143....9W} and derived emission line ratios ($[\ion{O}{iii}]$/H$\alpha$, $[\ion{S}{ii}]$/H$\alpha$, and $[\ion{O}{iii}]$/[\ion{S}{ii}]; see Appendix~\ref{Comparison of line ratios with other possible structures}) for a consideration of these alternative scenarios.

While the structure is bright in H$\alpha$ and $[\ion{S}{ii}]$, the low levels of $[\ion{O}{iii}]$ emission (see Appendix~\ref{Comparison of line ratios with other possible structures} and Fig.~\ref{OIII ratio}) and lack of radio emission (see MeerKAT panel in Fig.~\ref{observations}) from the shell (both typical of SNR shock fronts), the absence of radio sources within the structure, the low expansion velocities inferred from $\ion{\text{H}}{i}$ data (see Sect.~\ref{NSR in radio data}), and its size exceeding that of any SNR in the LMC \citep[e.g.,][]{2024A&A...692A.237Z} together demonstrate that the discovered structure is not an SNR. The absence of strong $[\ion{O}{iii}]$ emission (see Fig.~\ref{OIII ratio}) and the size of the shell \citep[LMC PNe are unresolved in MCELS data as they are arcsec-scale; e.g.,][]{2001ApJ...548..727S} rule out a PN origin. There is one young stellar object \citep{2008AJ....136...18W}, one cluster of stars \citep{2008MNRAS.389..678B}, and two associations of stars \citep{2008MNRAS.389..678B} within the shell. However, their classification does not specify whether they are OB associations \citep{2008MNRAS.389..678B}, which could ionise the local ISM and create a bubble of this size \citep{2020NewAR..9001549W}. Also, the structure is absent from the $\ion{\text{H}}{ii}$ region catalogue by \citet{2012ApJ...755...40P}, arguing against it being an $\ion{\text{H}}{ii}$ region or a superbubble. This is further supported by the lack of radio emission and the relatively high $[\ion{S}{ii}]$/H$\alpha$ ratio (see Fig.~\ref{SII ratio}).

\subsection{Confirmation of recurrent nature}\label{Confirmation of recurrent nature}
LMCN 1971-08a is deemed an RN on account of its spatial coincidence with the location of nova N LMC 1971b \citep{2016ApJ...818..145B}. Nevertheless, some scepticism has remained regarding its recurrent nature; for example, the entry for LMC 2009-02 (another identifier for LMCN 1971-08a) in Table 7 of \citet{2021MNRAS.505.3113O} is labelled as "perhaps" an RN, unlike other confirmed RNe in the LMC. Thus, the discovery of an NSR around this nova further supports its classification as an observationally defined RN, albeit from a different approach. In addition, owing to the existence of such an NSR, LMCN 1971-08a may have a much shorter recurrence period than the 38 years currently presumed and may therefore be observed in eruption much sooner than 2047.

\section{Summary}\label{Summary}
In this Letter, we detailed the discovery of a nova super-remnant in the LMC, hosted by the recurrent nova LMCN 1971-08a. The structure is spatially coincident with the nova's location, spans ${\sim}$200 pc in diameter -- making it the largest NSR found to date -- and is visible in narrowband H$\alpha$ and $[\ion{S}{ii}]$ emission indicative of shock-ionisation, but exhibits negligible $[\ion{O}{iii}]$ emission. A corresponding structure is apparent in $\ion{\text{H}}{i}$ data. Adding to the growing collection of such NSRs, this is the first to be discovered in the LMC, the brightest NSR observed to date, and only the second extragalactic nova shell identified after the M31 NSR.

\begin{acknowledgements}
We thank the referee for feedback that improved our study. MS and FZ acknowledges support from the DFG through the grants SA 2131/13-2, SA 2131/14-2, and SA 2131/15-2. This work made use of the high performance computing facilities at Liverpool John Moores University, partly funded by LJMU’s FET and by the Royal Society.

This project used data obtained with the Dark Energy Camera (DECam), which was constructed by the Dark Energy Survey (DES) collaboration. Funding for the DES Projects has been provided by the DOE and NSF (USA), MISE (Spain), STFC (UK), HEFCE (UK), NCSA (UIUC), KICP (U. Chicago), CCAPP (Ohio State), MIFPA (Texas A\&M), CNPQ, FAPERJ, FINEP (Brazil), MINECO (Spain), DFG (Germany) and the Collaborating Institutions in the Dark Energy Survey, which are Argonne Lab, UC Santa Cruz, University of Cambridge, CIEMAT-Madrid, University of Chicago, University College London, DES-Brazil Consortium, University of Edinburgh, ETH Zürich, Fermilab, University of Illinois, ICE (IEEC-CSIC), IFAE Barcelona, Lawrence Berkeley Lab, LMU München and the associated Excellence Cluster Universe, University of Michigan, NSF NOIRLab, University of Nottingham, Ohio State University, OzDES Membership Consortium, University of Pennsylvania, University of Portsmouth, SLAC National Lab, Stanford University, University of Sussex, and Texas A\&M University.

Based on observations at NSF Cerro Tololo Inter-American Observatory, NSF NOIRLab (NOIRLab Prop. ID 2018B-0908; PI: T. Puzia) and (NOIRLab Prop. ID 2021B-0060; PI: S. Points), which is managed by the Association of Universities for Research in Astronomy (AURA) under a cooperative agreement with the U.S. National Science Foundation.

This research has made use of the SIMBAD database, operated at CDS, Strasbourg, France \citep[see][]{2000A&AS..143....9W}. This research has made use of the VizieR catalogue access tool, CDS, Strasbourg, France. The original description of the VizieR service was published in \citet{2000A&AS..143...23O}.
\end{acknowledgements}

\bibliographystyle{aa}
\bibliography{bibliography.bib}

\begin{appendix}

\section{Further remarks}
\subsection{NSR missed in previous work}\label{Missed in previous work}
The LMC was surveyed in previous work \citep{2024MNRAS.528.3531H} through a targeted search around the four recurrent novae using the Faulkes Telescope South but no NSRs were uncovered. While $1.5\times1.5$ arcminute FOV images were provided in \citet{2024MNRAS.528.3531H} around LMCN 1971-08a, the field of view of the Faulkes Telescope South facility used was $10.5\times10.5$ arcminutes (chosen to search the surroundings of the RNe for this reason by \citealt{2024MNRAS.528.3531H}). As such, the region observed was large enough to capture a component of the NSR, however the relatively shorter exposure time of the observations ($3 \times 720$s for H$\alpha$) and therefore fainter emission, in addition to (and more importantly) only a component of the NSR being captured, indicates why this structure was overlooked in the previous work. The FOV of the Faulkes Telescope South observation from \citet{2024MNRAS.528.3531H} is shown in Fig.~\ref{FTS}. Now with knowledge of the gross structure of the NSR, the southwest shell edge can be made out when comparing with the DeMCELS image in Fig.~\ref{FTS}, as can the structure to the west of the nova's position.

\subsection{Surroundings of other LMC recurrent novae}\label{Other LMC recurrent novae}
The surroundings of all four RNe in the LMC were initially checked within the H$\alpha$ data of MCELS for signs of a NSR. While the NSR was discovered around LMCN 1971-08a, little is seen in the environments around LMCN 1968-12a, LMCN 1996 and YY Doradus. 

\subsection{Comparison of line ratios with other possible structures}\label{Comparison of line ratios with other possible structures}
The ratios of emission lines can be used as diagnostic tools for helping to differentiate between different extended structures. As such, in Fig.~\ref{OIII ratio}, we compare the $[\ion{O}{iii}]/\text{H}\alpha$ ratio for the nova super-remnant with the same ratio for three LMC supernova remnants, J0450-7050 \citep[a.k.a. Veliki;][]{Veliki2025} and J0534-7033 and J0506-6541 \citep[from][]{2024A&A...692A.237Z}; the LMC $\ion{\text{H}}{ii}$ region, L77 DEM L45 \citep[from][]{2012ApJ...755...40P}; and the LMC planetary nebula, SMP LMC 80 \citep[from][]{2024A&A...688A..36T}, in order to differentiate the shell from other structures. We also provide similar comparative plots for the ratios $[\ion{S}{ii}]/\text{H}\alpha$ and $[\ion{O}{iii}]/[\ion{S}{ii}]$ in Figs.~\ref{SII ratio} and~\ref{OIII/SII ratio}, respectively.

The ratios were calculated for a number of apertures (with radii of $5''$) along a slit passing through the structure (illustrated in Figs.~\ref{OIII ratio}, ~\ref{SII ratio}, and \ref{OIII/SII ratio}). As the structures are different sizes, the length of the line (and therefore the number of apertures used) were scaled accordingly such that the ratios can be compared. The slit shown in Fig.~\ref{OIII ratio} is ${\sim}1000''$ long with 100 apertures for the nova super-remnant; ${\sim}40''$ long with 4 apertures for planetary nebula SMP LMC 80 (as PNe in the LMC are compact); ${\sim}810''$ long with 80 apertures for $\ion{\text{H}}{ii}$ region L77 DEM L45; ${\sim}400''$ long with 40 apertures for SNR J0450-7050; ${\sim}260''$ long with 25 apertures for SNR J0534-7033; and ${\sim}530''$ long with 50 apertures for SNR J0506-6541. In the ratio plots of Figs. \ref{OIII ratio}, \ref{SII ratio}, and \ref{OIII/SII ratio}, the $x$-axes have been normalised.

\paragraph{Supernova remnants}
The emission line ratio $[\ion{S}{ii}]/\text{H}\alpha$ for supernova remnants is typically high \citep[$>0.4$ and see, e.g.,][]{2021MNRAS.500.2336Y,2025ApJ...983..150C} as a result of enhanced ionisation and excitation of sulphur from radiative shocks \citep{2024A&A...692A.237Z}. In addition to this, SNRs also exhibit strong $[\ion{O}{iii}]$ emission as a result of shock heating \citep{2021MNRAS.500.2336Y}. While the nova super-remnant does exhibit a $[\ion{S}{ii}]/\text{H}\alpha$ ratio on par with the example supernova remnants ($\log([\ion{S}{ii}]/\text{H}\alpha)\sim-0.1$ and see Fig.~\ref{SII ratio}), there is a large difference in the $[\ion{O}{iii}]/\text{H}\alpha$ ratios for the NSR and the SNRs, with the NSR $[\ion{O}{iii}]$ emission being much fainter ($\log([\ion{O}{iii}]/\text{H}\alpha)<-0.5$) within the NE and SW component of the shell (between ${\sim}0.05-0.13$ and ${\sim}0.86-0.95$ on the $x-$axis of Fig.~\ref{OIII ratio}).

\paragraph{Planetary nebulae}
Planetary nebulae have bright $[\ion{O}{iii}]$ emission \citep[][and see the example PN in Fig.~\ref{OIII ratio}]{2013A&A...557A.121G} as a result of fast outflows shocking the surrounding low-density material \citep{2013A&A...557A.121G} and weak $[\ion{S}{ii}]$ emission \citep[see, e.g.,][]{2025ApJ...983..150C}. However, as can be seen in Fig.~\ref{OIII ratio}, the nova super-remnant is faint in $[\ion{O}{iii}]$ (with $\log([\ion{O}{iii}]/\text{H}\alpha)<-0.5$ in the NE and SW component of the shell) and has moderate levels of $[\ion{S}{ii}]$ emission (see Fig.~\ref{SII ratio}).

\paragraph{$\ion{\text{H}}{ii}$ regions}
Typical $\ion{\text{H}}{ii}$ regions have low $[\ion{S}{ii}]/\text{H}\alpha$ ratios as a result of the sulphur in such structures existing primarily as doubly-ionised sulphur \citep{2021MNRAS.500.2336Y} from the gas being photoionised \citep{2024A&A...692A.237Z}. As can be seen in Fig.~\ref{SII ratio}, the nova super-remnant has a higher $[\ion{S}{ii}]/\text{H}\alpha$ ratio across the shell than the ratio of a typical $\ion{\text{H}}{ii}$ region. In addition, as illustrated with the example $\ion{\text{H}}{ii}$ region in Fig.~\ref{OIII ratio}, $\ion{\text{H}}{ii}$ regions have high levels of $[\ion{O}{iii}]$ due to photoionisation from hot central stars \citep{2017PASP..129h2001P} however the discovered structure does not with $\log([\ion{O}{iii}]/\text{H}\alpha)<-0.5$ for the NSR compared to $\log([\ion{O}{iii}]/\text{H}\alpha)>0.2$ for the $\ion{\text{H}}{ii}$ region.

\onecolumn
\section{Figures}
\begin{figure*}[h!]
\centering
\includegraphics[width=0.8\textwidth]{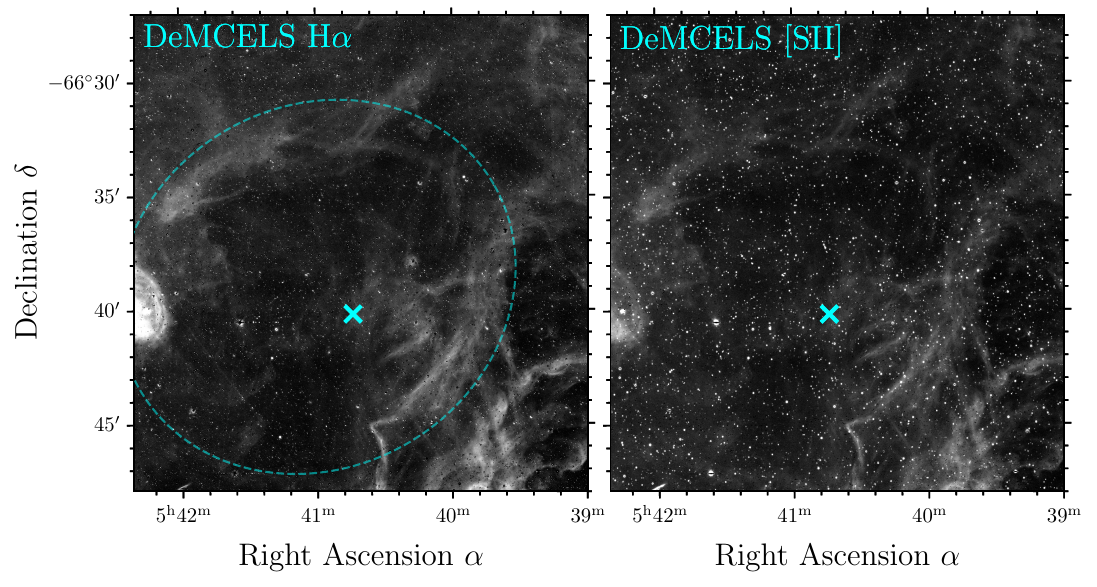}
\caption{Left: DeMCELS continuum-subtracted H$\alpha$ image of the nova super-remnant surrounding the recurrent nova LMCN 1971-08a. The dashed ellipse is provided to approximately indicate the location of the NSR and its extent. The location of the nova is indicated by the cyan cross. Right: DeMCELS continuum-subtracted $[\ion{S}{ii}]$ image of the same region. As with Fig.~\ref{observations}, the bright "stars" in the DeMCELS images are points of oversubtraction resulting from saturated stars in the R-band data used for continuum subtraction.}
\label{DeMCELS}
\end{figure*}

\begin{figure*}[h!]
\centering
\includegraphics[width=0.62\textwidth]{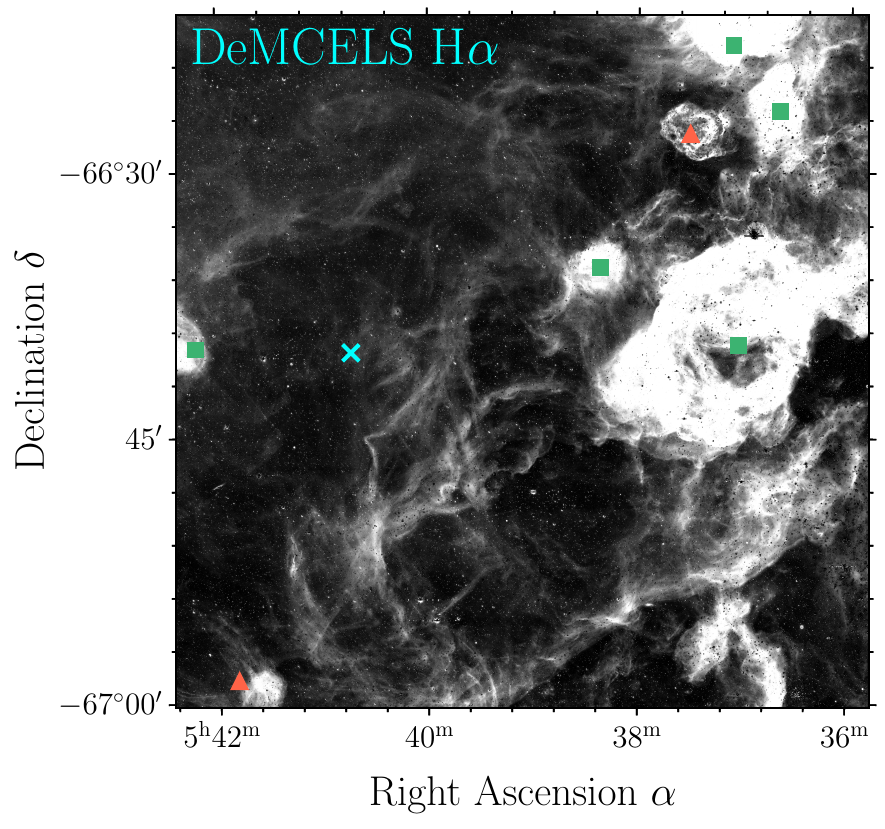}
\caption{DeMCELS continuum-subtracted H$\alpha$ image with an approximately $39 \times 39$ arcminute field of view, showing the nova super-remnant surrounding LMCN 1971-08a and the extended environment including the multitude of $\ion{\text{H}}{ii}$ regions to the west. The location of the nova is indicated by the cyan cross near the centre of the NSR shell. Other catalogued structures in the vicinity are labelled as follows: $\ion{\text{H}}{ii}$ regions (green square) from \citet{2012ApJ...755...40P} and SNRs (red triangles) from \citet{2024A&A...692A.237Z}.}
\label{full DeMCELS}
\end{figure*}

\begin{figure*}[h!]
\centering
\includegraphics[width=0.33\textwidth]{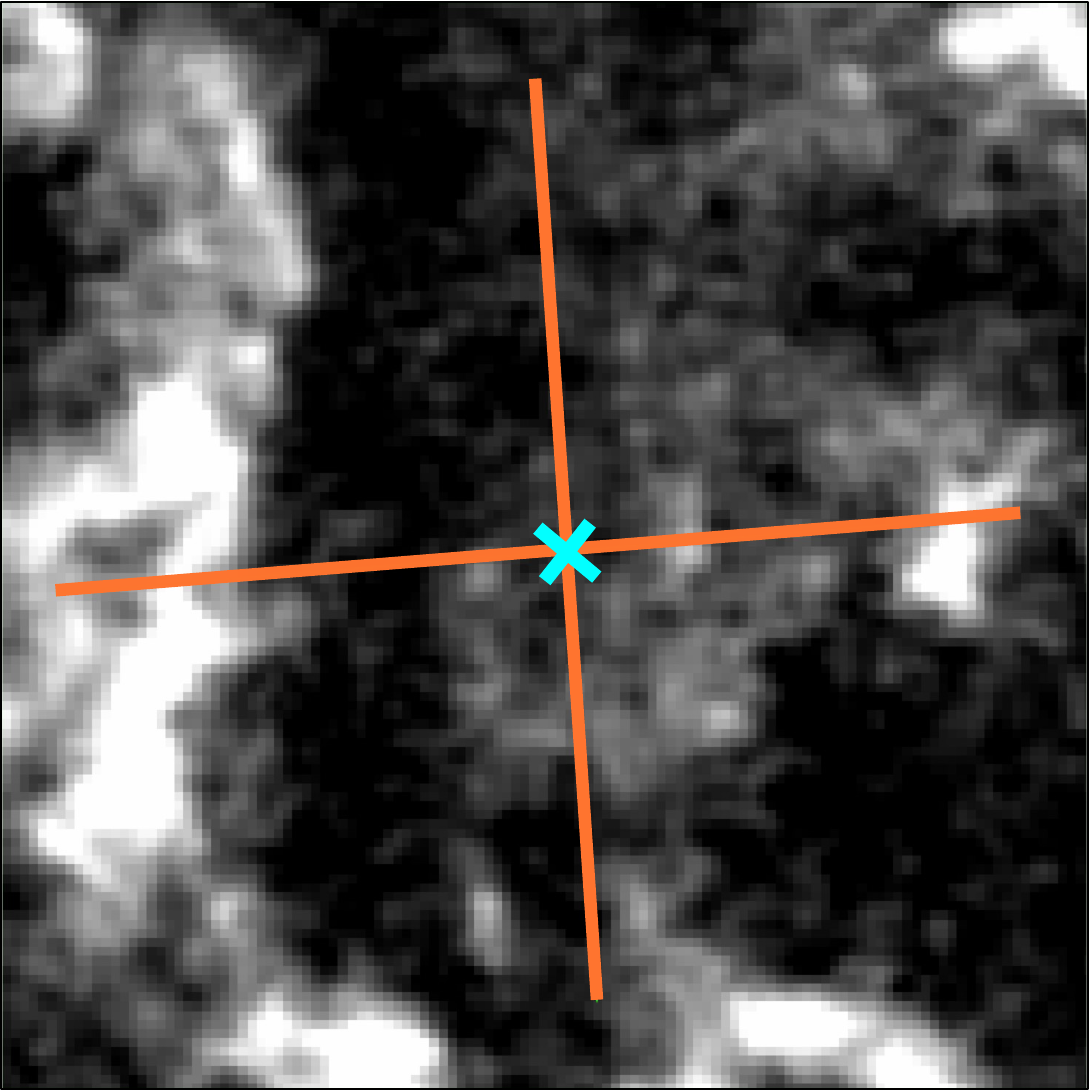}
\caption{HI4PI survey image, as shown in Fig.~\ref{observations} (at ${\sim}294 \ \text{km} \ \text{s}^{-1}$), with the two slits used for the position-velocity analysis indicated with orange lines (one at constant right ascension and the other at constant declination). The location of the nova is marked with a cyan cross.}
\label{HI slits}
\end{figure*}

\begin{figure*}[h!]
\centering
\includegraphics[width=0.46\linewidth]{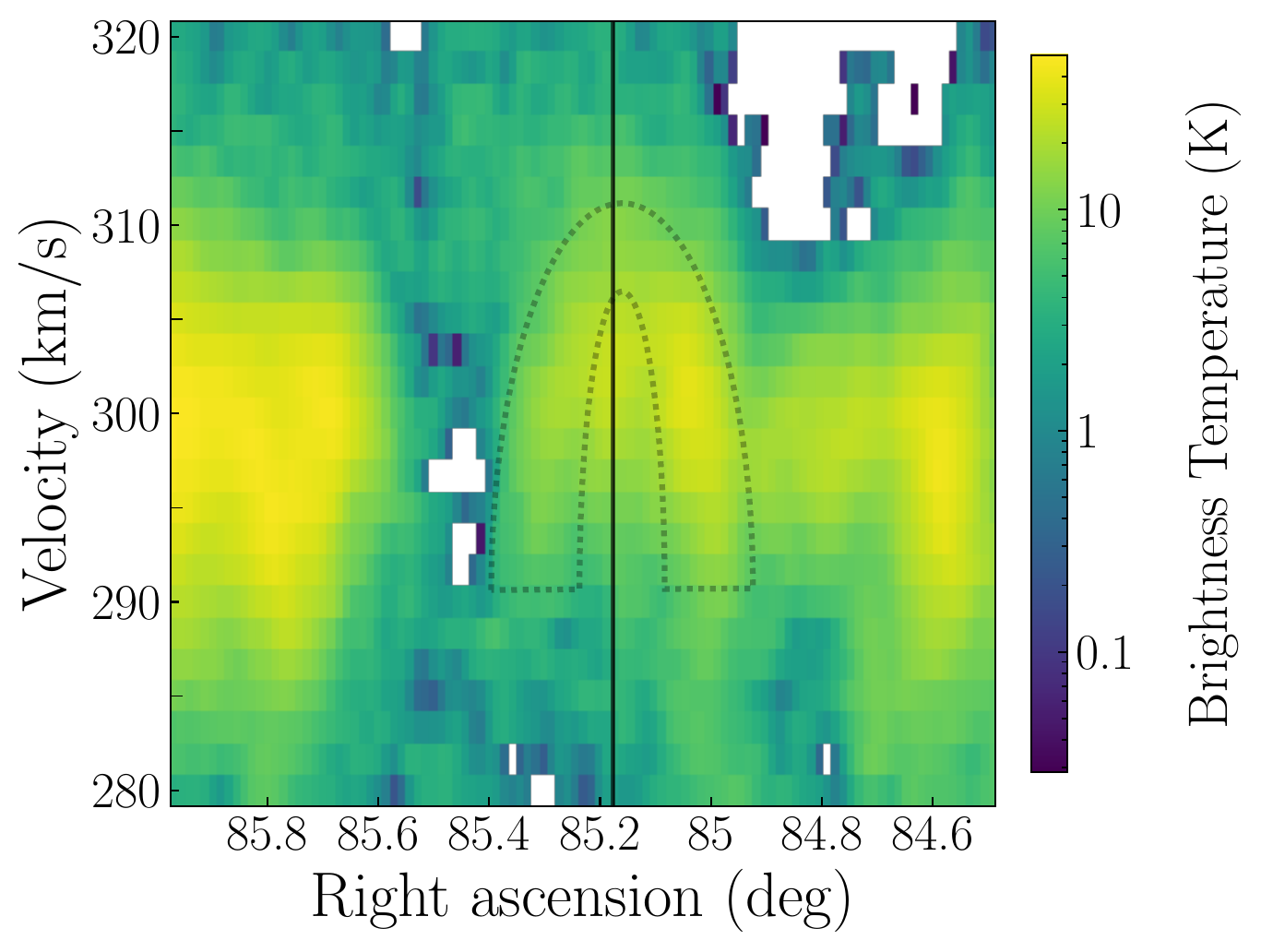}
\includegraphics[width=0.46\linewidth]{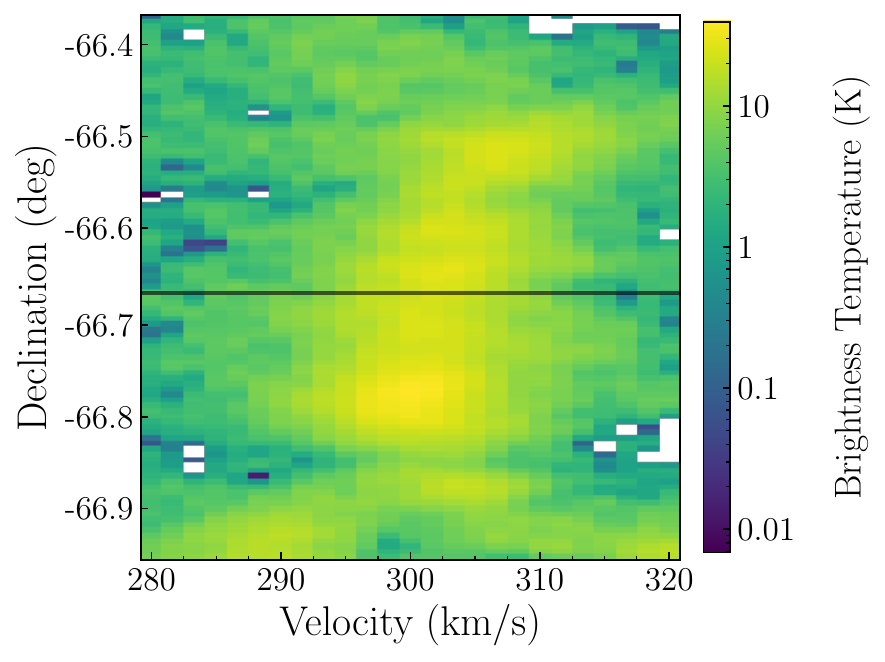}
\caption{Parkes $\ion{\text{H}}{i}$ position-velocity diagrams around the position of the nova super-remnant. The location of LMCN 1971-08a is indicated by the dashed line. The arc ranging from ${\sim}290-310 \ \text{km} \ \text{s}^{-1}$ is approximately outlined with a dashed line.}
\label{PV diagram}
\end{figure*}

\begin{figure*}[h!]
\centering
\includegraphics[width=0.64\columnwidth]{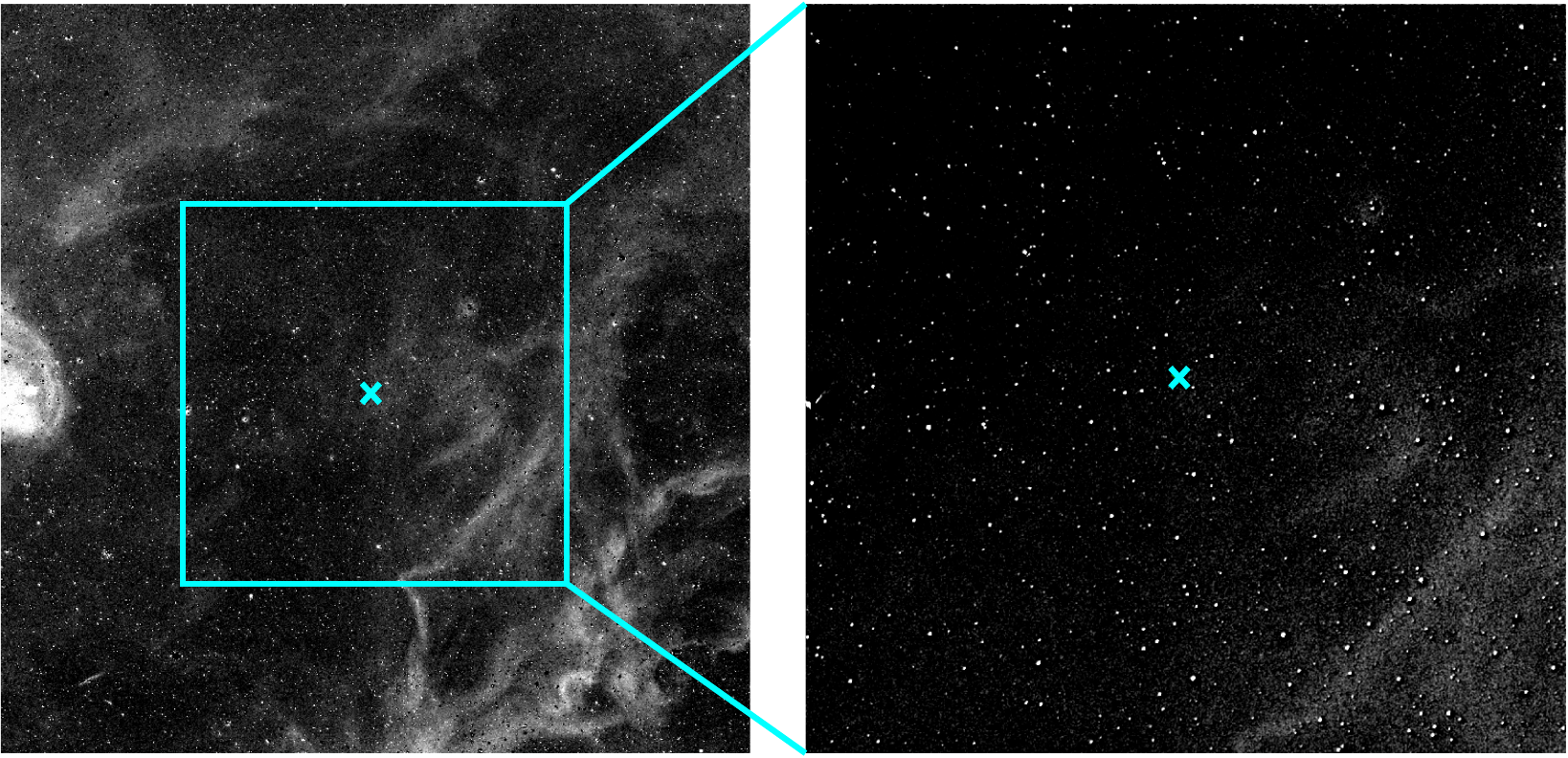}
\caption{Faulkes Telescope South $10.5 \times 10.5$ arcminute field of view with the H$\alpha$ filter used in \citet{2024MNRAS.528.3531H} overlaid onto the H$\alpha$ DeMCELS data (left panel). A part of the NSR was detected but overlooked in previous studies, mainly because the larger structure remained unidentified, and it was faint. The location of the nova is marked with the cyan cross.}
\label{FTS}
\end{figure*}

\begin{figure*}[h!]
\centering
\includegraphics[width=\columnwidth]{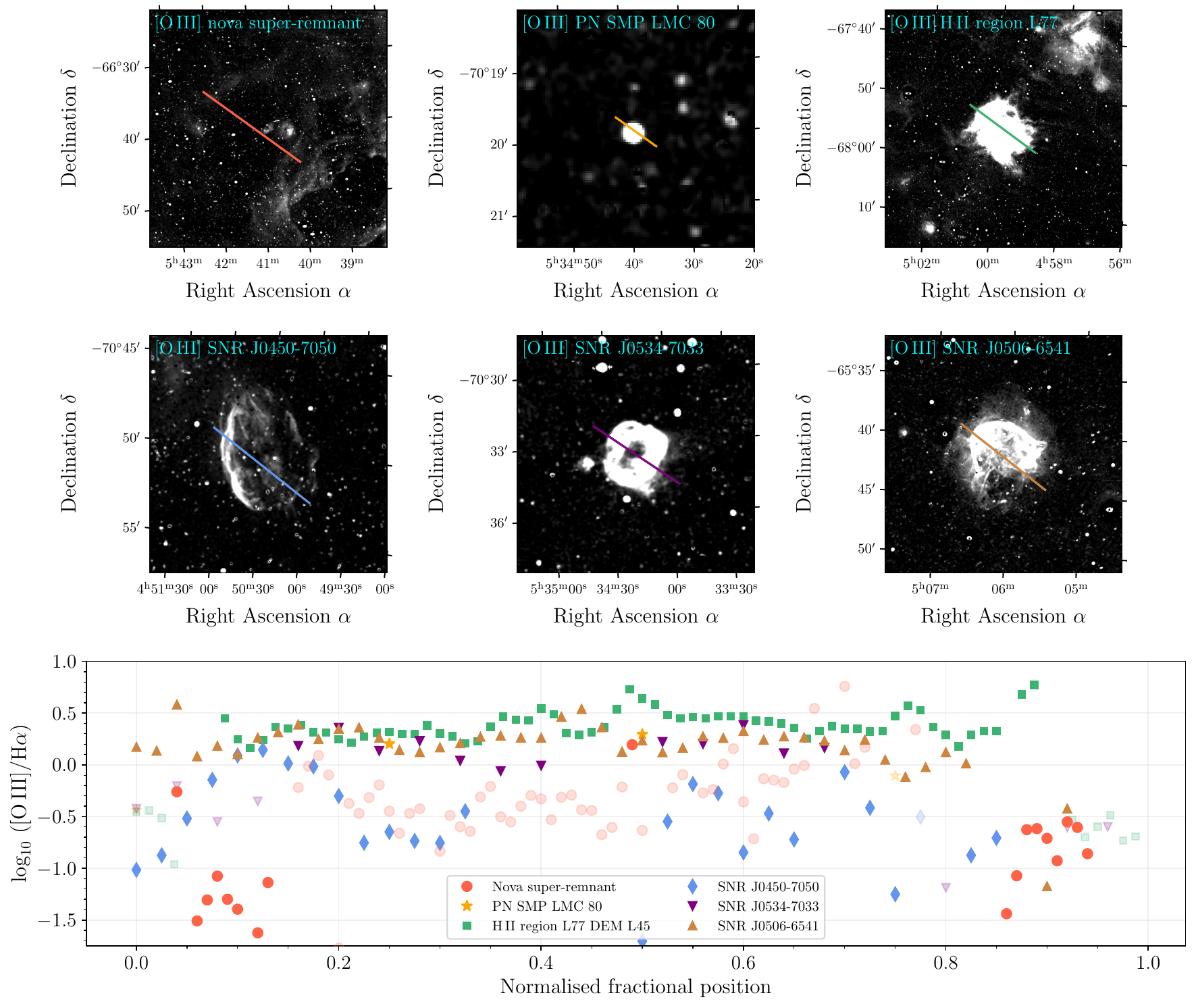}
\caption{$[\ion{O}{iii}]/\text{H}\alpha$ ratios for the nova super-remnant, three supernova remnants, an $\ion{\text{H}}{ii}$ region, and a planetary nebula. Top and middle rows: MCELS continuum-subtracted  $[\ion{O}{iii}]$ images of the six sources, each with a different field of view, as follows: ${\sim}33'\times33'$ for the NSR; ${\sim}3'\times3'$ for PN SMP LMC 80; ${\sim}40'\times40'$ for $\ion{\text{H}}{ii}$ region L77 DEM L45; ${\sim}13'\times13'$ for SNR J0450-7050 \citep[a.k.a. Veliki;][]{Veliki2025}; ${\sim}10'\times10'$ for SNR J0534-7033; and ${\sim}20'\times20'$ for SNR J0506-6541. As detailed in Appendix~\ref{Comparison of line ratios with other possible structures}, slits with different lengths were placed across each structure (with different numbers of apertures) to derive ratios. These slits are shown in each panel. Bottom row: Comparison of the $[\ion{O}{iii}]/\text{H}\alpha$ ratios for the six sources across the length of the slit with the $x-$axis normalised to the length of the slit. The nova super-remnant has two distinct groupings of ratios (between ${\sim}0.05-0.13$ and ${\sim}0.86-0.95$) corresponding to the northeast and southwest shell with low $[\ion{O}{iii}]/\text{H}\alpha$ ratios. Two of the supernova remnants (J0534-7033 and J0506-6541) and the $\ion{\text{H}}{ii}$ region have a high $[\ion{O}{iii}]/\text{H}\alpha$ ratio across their full extent, while the SNR J0450-7050 has a high $[\ion{O}{iii}]/\text{H}\alpha$ across the eastern edge. The $[\ion{O}{iii}]/\text{H}\alpha$ ratios from the two apertures across the PN are also high, as expected. The transparent points are from apertures with flux sums $\le 0$.}
\label{OIII ratio}
\end{figure*}

\begin{figure*}[h!]
\centering
\includegraphics[width=\columnwidth]{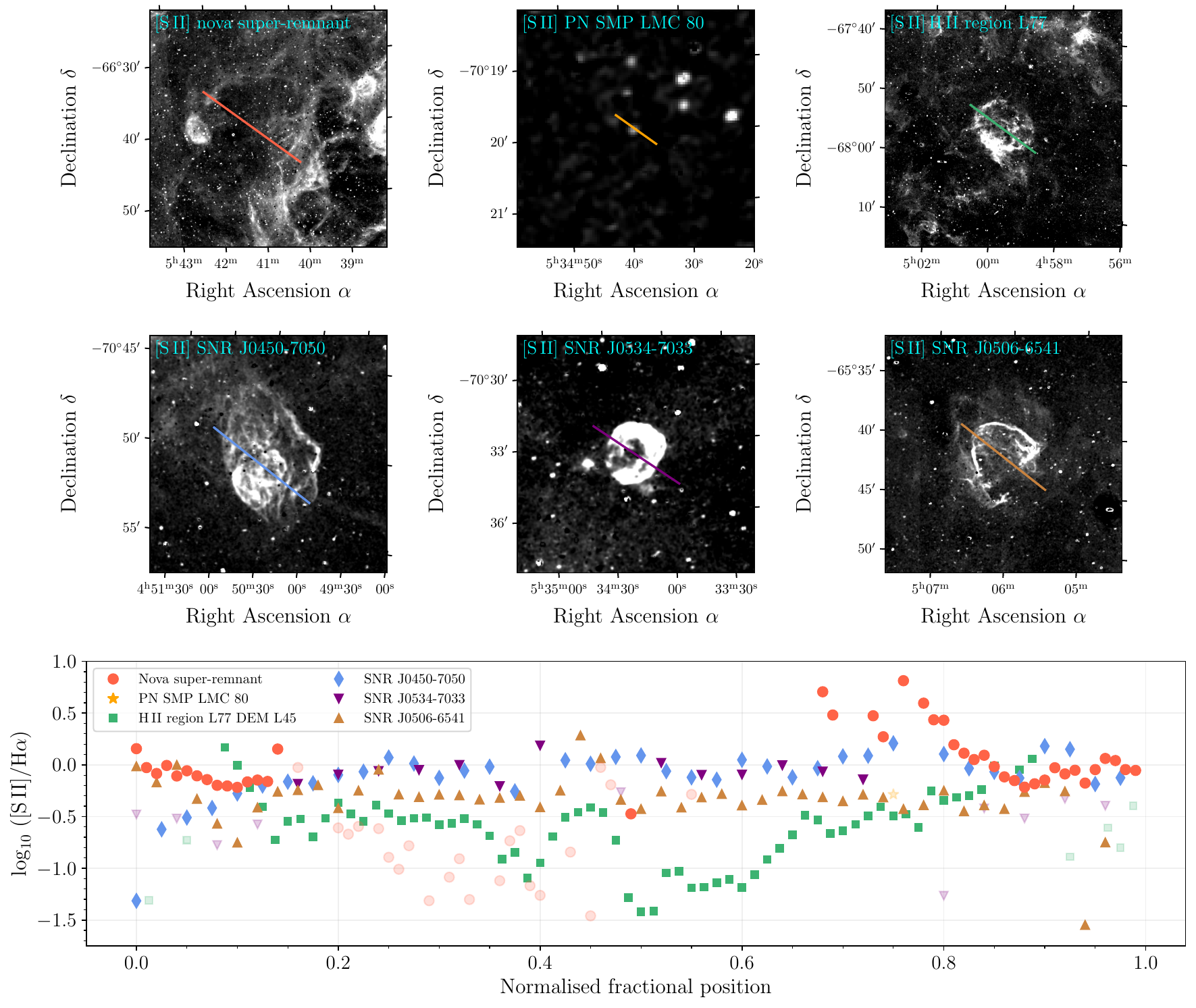}
\caption{Same as Fig.~\ref{OIII ratio} but for the ratio $[\ion{S}{ii}]/\text{H}\alpha$. The top six panels show the MCELS continuum-subtracted $[\ion{S}{ii}]$ images for the same six structures as in Fig.~\ref{OIII ratio}. The bottom panel shows a comparison of the $[\ion{S}{ii}]/\text{H}\alpha$ ratio across the structures.}
\label{SII ratio}
\end{figure*}

\begin{figure*}[h!]
\centering
\includegraphics[width=\columnwidth]{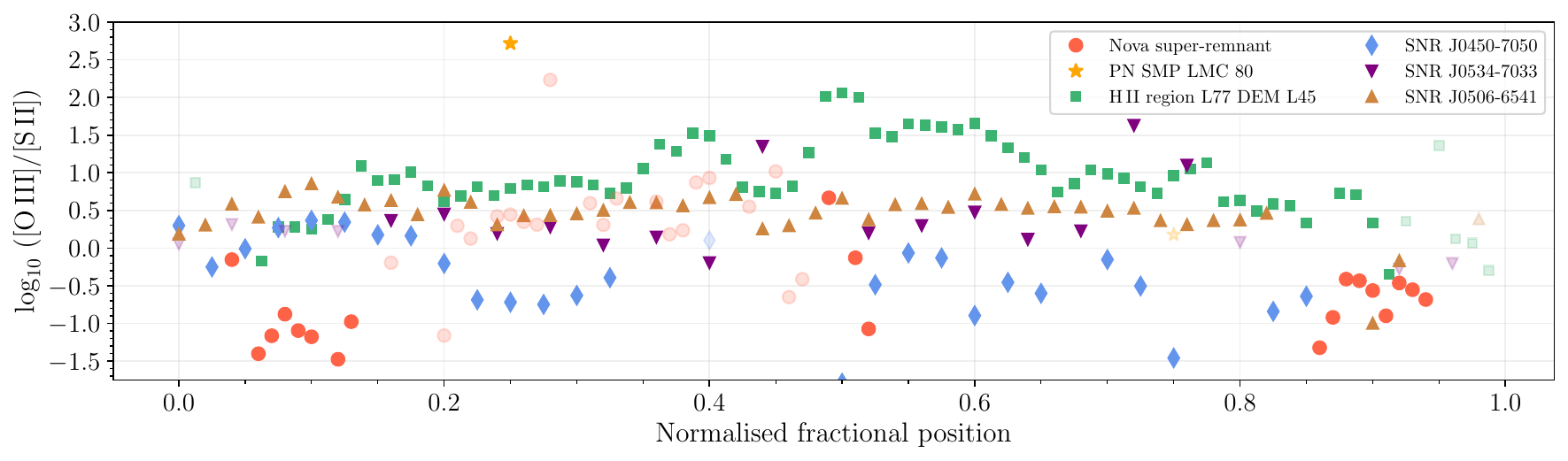}
\caption{Comparison of the $[\ion{O}{iii}]/[\ion{S}{ii}]$ ratio across the structure for the six sources shown in Fig.~\ref{OIII ratio}. The line of apertures used is the same as in Fig.~\ref{OIII ratio}. Note that the $y$-axis scale ranges from $-1.75$ to 3 here.}
\label{OIII/SII ratio}
\end{figure*}

\end{appendix}
\end{document}